\documentclass[11pt]{article}
\usepackage{anysize}
\marginsize{2cm}{2cm}{2cm}{2cm}
\usepackage{amsmath,amsfonts,amssymb,amsthm,verbatim,color,lscape}
\usepackage[figuresright]{rotating}
\usepackage[super,comma,sort&compress]{natbib}

\usepackage{url}
\usepackage{hyperref}
\hypersetup{colorlinks,
citecolor=black,
filecolor=black,
linkcolor=black,
urlcolor=black,
pdftex}

\usepackage{multicol}

\usepackage[font={footnotesize}]{caption}

\usepackage{sectsty}  
\subsectionfont{\normalsize\bf}
\sectionfont{\large\bf}

\def\Cbar{{\overline C}}

\def\R{\mathbf{R}}

\def\x{\mathbf{x}}

\def\s{\sigma}

\def\X{\mathbf{X}}

\def\a{\alpha}
\def\g{\gamma}
\def\b{\beta}

\def\th{\theta}
\def\thbf{\boldsymbol{\theta}}

\def\Sbf{\boldsymbol{\Sigma}}

\def\bmu{\boldsymbol{\mu}}

\long\def\symbolfootnote[#1]#2{\begingroup
\def\thefootnote{\fnsymbol{footnote}}\footnote[#1]{#2}\endgroup}

\newcommand{\widesim}[2][1.5]{
  \mathrel{\overset{#2}{\scalebox{#1}[1]{$\sim$}}}
}
\begin{document}

\title{A vine copula mixed effect model  
for trivariate \\meta-analysis of diagnostic test accuracy studies \\accounting for disease prevalence}

\author{Aristidis~K.~Nikoloulopoulos\footnote{{\small\texttt{A.Nikoloulopoulos@uea.ac.uk}}, School of Computing Sciences, University of East Anglia,
Norwich Research Park, Norwich NR4 7TJ, UK}}
\date{}

\maketitle

\begin{abstract}
\baselineskip=19pt
\noindent 
A bivariate copula mixed model  has been recently proposed   to synthesize
diagnostic test accuracy studies and it has been shown that is superior to the standard generalized linear mixed model (GLMM) in this context.
Here we call trivariate vine copulas to extend the bivariate meta-analysis of diagnostic test accuracy studies by accounting for disease prevalence. 
Our vine copula mixed model includes the trivariate GLMM as a special case and can also operate on the original scale of sensitivity, specificity, and disease prevalence.
Our general
methodology is  illustrated by re-analysing the data of two  published meta-analyses. 
Our study suggests that
there can be an improvement on trivariate GLMM in fit to data and makes the argument for moving to vine copula random effects models especially because of their richness including reflection asymmetric tail dependence,  and, computational feasibility despite their three-dimensionality. 
\medskip

\noindent \textbf{Key Words:} copula models; diagnostic tests; multivariate meta-analysis;  random effects models;  sensitivity/specificity/prevalence; vines. 
\end{abstract}

\maketitle

\baselineskip=19pt

\section{Introduction}
Synthesis of diagnostic test accuracy studies is  the most common medical application of multivariate meta-analysis.\cite{JacksonRileyWhite2011,
MavridisSalanti13,Ma-etal-2013}
Diagnostic test accuracy studies observe the result of a gold standard procedure which defines the presence or absence of a disease and the result of  a diagnostic test.  They
typically report the number of true positives (diseased
people correctly diagnosed), false positives (non-diseased people incorrectly diagnosed as diseased),
true negatives and false negatives.

In situations where studies compare a diagnostic test with its gold standard,
heterogeneity arises between studies due to the differences in disease prevalence,
study design as well as laboratory and other characteristics.\cite{Chu-etal-2012} Because of this heterogeneity,
a generalized linear mixed model (GLMM) 
has been recommended in the biostatistics literature  to synthesize information.\cite{Chu&Cole2006,Arends-etal-2008,hamza-etal-2009,Ma-etal-2013} 
The GLMM assumes independent binomial distributions for the true positives and true negatives, conditional on the latent pair of transformed (via a link function) sensitivity
and specificity  in each study. The  random effects (latent pair of transformed sensitivity
and specificity) are jointly analysed with a bivariate normal (BVN) distribution.

However, it is reported in the literature that sensitivity and specificity may not reflect the clinical utility of a diagnostic test; such clinical utility depends on the prevalence of disease in the population to which the diagnostic test is applied.\cite{chu-etal-2009} 
Furthermore, many empirical studies have shown the assumption of independence between the  sensitivity/specificity with disease prevalence for a dichotomous disease status is likely to be violated. 
\cite{brenner-gefeller-1997,leeflang-etal-2009,Leeflang-etal-2013}

To this end, Chu {\it et al.}\cite{chu-etal-2009} extended the bivariate GLMM to a trivariate GLMM by also accounting for disease prevalence.  
In the same manner, the trivariate GLMM assumes independent binomial distributions for the true positives, true negatives, and diseased persons conditional on a latent vector of transformed (via a link function) sensitivity, specificity, and prevalence  in each study. The  random effects herein are jointly analysed with a trivariate normal (TVN) distribution.

In this paper we propose a vine copula mixed model as an extension of the trivariate GLMM  by rather using a vine copula representation of the random effects distribution  with normal and  beta margins. Our general model (a) includes the trivariate GLMM as a special case, (b) can also operate on the original scale of sensitivity, specificity, and prevalence, and (c) can also provide tail dependencies and asymmetries.

In fact, we extend the copula mixed model proposed by Nikoloulopoulos \cite{nikoloulopoulos2015remada}  to the trivariate case accounting for disease prevalence.  There are many simple bivariate copula families, but generally
their multivariate extensions have limited dependence structures. For example, 
Archimedean copulas in more that two dimensions allow
only exchangeable structure with a narrower range of negative dependence
as the dimension increases.  \cite{joe97,mcneil&neslehova08}
The multivariate normal (MVN) copula generated by the MVN distribution inherits the useful properties of the latter, thus allowing a wide range for dependence, and overcomes the drawback of limited dependence inherent in simple parametric families of copulas.\cite{nikoloulopoulos&joe&chaganty10}  However,  it provides tail independence and reflection symmetry, thus it is not suitable when data display dependence among extreme values and inferences based on multivariate tail probabilities are needed.\cite{joeetal10} 
In Nikoloulopoulos\cite{nikoloulopoulos2015remada} it has shown  that such asymmetries are expected in binary 
data from diagnostic accuracy test studies and summary receiver operating characteristic  inference is particularly based on bivariate tail probabilities.

In recent years, a popular and useful approach is the vine pair-copula construction
which is based on $d(d-1)/2$ bivariate copulas, of which some are
used to summarize conditional dependence; a special case occurs if all of these
bivariate copulas are BVN, and then the parametrization of an MVN copula is a set of correlations and partial correlations that
are algebraically independent in $(-1,1)^{d(d-1)/2}$.\cite{Kurowicka-Joe-2011,joe2014} 
The $d$-dimensional vine copulas
can cover flexible dependence structures through the specification
of $d-1$ bivariate marginal copulas at level 1 and $(d-1)(d-2)/2$
bivariate conditional copulas at higher levels; at level $\ell$  for
$\ell=2,\ldots,d-1$, there are $d-\ell$ bivariate conditional copulas
that condition on $\ell-1$ variables.  Vine copulas include MVN as special case, but can also cover reflection
asymmetry and have upper/lower tail dependence parameters being different
for each bivariate margin $(i,j)$.  Joe {\it et al.}\cite{joeetal10} have a main theorem
that says that all bivariate margins of the vine copula have upper/lower
tail dependence if the bivariate copulas at level 1 have upper/lower
tail dependence.

A vine copula approach for meta-analysis of diagnostic accuracy studies was recently proposed by Hoyer and Kuss \cite{hoyer&kuss2015} who explored the use of  a trivariate vine copula model for observed discrete variables (number of true positives,  true negatives, and diseased persons) which have beta-binomial margins. This model is actually an approximation of the vine copula mixed model with beta margins for the latent vector of sensitivity, specificity, and prevalence.
Nikoloulopoulos \cite{nikoloulopoulos2015remada}  has  extensively studied the small-sample and theoretical efficiency  of this approximation in the bivariate case.\cite{kuss-etal-2013} It was clearly shown that this approximation is an  inefficient approach;  it leads  to substantial downward bias for the estimates  of the dependence and bias for the meta-analytic parameters for fully specified copula mixed models.  This evolves because there are serious problems on modelling assumptions under the case of heterogeneous  study sizes. If
the number of true positives, true negatives, and diseased persons  do not have a common support over different studies,  then one cannot conclude that there is a copula.

Furthermore, Hoyer and Kuss \cite{hoyer&kuss2015}
compared fits and inference for vine copulas, when the bivariate copulas
are all reflection symmetric, i.e., Placket or BVN. 
In this paper, we make the first use of vine copulas with 
bivariate linking copulas that can have upper tail dependence different
from lower tail dependence.  This provides a means to check if there is
some reflection asymmetry in the joint tails of  diagnostic accuracy data from binary test
outcomes.

The remainder of the paper proceeds as follows. Section \ref{stand-model-sec} summarizes the standard GLMM for synthesis of diagnostic test accuracy studies accounting for disease prevalence.
Section \ref{copula-mixed-model-sec} has a brief overview of
relevant vine copula theory and then introduces the vine copula mixed model for diagnostic test accuracy studies accounting for disease prevalence, discusses its relationship with existing  models, and provides computational details for maximum likelihood estimation.
 Section \ref{construction} discusses  practical issues for (trivariate) vine copula modelling that have been neglected in Hoyer and Kuss \cite{hoyer&kuss2015}. Section \ref{miss-section} contains  small-sample  efficiency calculations
to  investigate the effect of misspecifying the random effects distribution on parameter estimators and standard errors. 
Section \ref{sec-appl} presents applications of our methodology to two data frames with diagnostic accuracy data from binary test
outcomes. We conclude with some discussion in Section \ref{sec-discussion}, followed by a brief section with the software details.

\section{\label{stand-model-sec} The standard GLMM }

We first introduce the notation used in this paper. The focus is on two-level (within-study and between-studies) cluster data. The data  are $(y_{ij}, n_{ij}),\, i = 1, . . . ,N,\, j=1,2,3$, where $j$ is an index for the within study measurements and $i$ is an
index for the individual studies.

The standard two-level model of meta-analysing diagnostic test accuracy studies accounting for disease prevalence\cite{chu-etal-2009,Ma-etal-2013}  lies in the framework of mixed models \cite{Demidenko04}.
The within-study model assumes that the number of true positives $Y_{i1}$, true negatives $Y_{i2}$, and diseased persons $Y_{i3}$ are conditionally independent and binomially distributed given $\X=\x$, where $\X=(X_1,X_2,X_3)$ denotes the  trivariate latent (random) vector of transformed sensitivity, specificity, and disease prevalence.  That is
\begin{eqnarray}\label{withinBinom}
Y_{i1}|X_{1}=x_1&\sim& \mbox{Binomial}\Bigl(n_{i1},l^{-1}(x_1)\Bigr);\nonumber\\
Y_{i2}|X_{2}=x_2&\sim& \mbox{Binomial}\Bigl(n_{i2},l^{-1}(x_2)\Bigr);\\
Y_{i3}|X_{3}=x_3&\sim& \mbox{Binomial}\Bigl(n_{i3},l^{-1}(x_3)\Bigr),\nonumber
\end{eqnarray}
where $l(\cdot)$ is a link function such as the commonly used logit.
The between studies model assumes
that $\X$ is TVN distributed with mean vector $\bmu=\bigl(l(\pi_1),l(\pi_2),l(\pi_3)\bigr)^\top$ and variance covariance matrix
$\Sbf=\begin{pmatrix}
\sigma_1^2 &\rho_{12}\sigma_1\s_2 &\rho_{13}\sigma_1\s_3\\
\rho_{12}\sigma_1\sigma_2 & \sigma_2^2&\rho_{23}\sigma_2\s_3\\
\rho_{13}\sigma_1\sigma_3 &\rho_{23}\sigma_2\s_3& \sigma_3^2
\end{pmatrix}$. That is
\begin{equation}\label{between}
\X
\sim
\mbox{TVN}
\bigl(\bmu,\Sbf\bigr).
\end{equation}
The models in (\ref{withinBinom}) and (\ref{between}) together specify a GLMM with joint likelihood
$$
L(\pi_1,\pi_2,\pi_3,\sigma_1,\sigma_2,\sigma_3,\rho_{12},\rho_{13},\rho_{23})=\prod_{i=1}^N\int\int
\prod_{j=1}^3g\Bigl(y_{ij};n_{ij},l^{-1}(x_j)\Bigr)\phi_{123}(x_1,x_2,x_3;\bmu,\Sbf)dx_1dx_2dx_3,
$$
where
$$g\bigl(y;n,\pi\bigr)=\binom{n}{y}\pi^y(1-\pi)^{n-y},\quad y=0,1,\ldots,n,\quad 0<\pi<1,$$
 is the binomial probability mass function (pmf) and $\phi_{123}(\cdot;\bmu,\Sbf)$ is the TVN density with mean vector $\bmu$  and variance covariance matrix $\Sbf$.
The parameters $\pi_1$, $\pi_2$ and $\pi_3$ are those of actual interest denoting the meta-analytic parameters for the sensitivity, specificity, and disease prevalence, respectively, while the univariate parameters $\s_1^2$, $\s^2_2$, and $\s^2_3$  are of secondary interest denoting the variability between studies.

\section{\label{copula-mixed-model-sec}The vine copula mixed model for diagnostic test accuracy studies  }
In this section, we  introduce the vine copula mixed model for diagnostic test accuracy studies and discuss its relationship with existing  models. Before that, the first subsection has some background on copula models.
In Subsection \ref{normal-parametrization} and Subsection \ref{beta-parametrization} a vine copula representation of the random effects distribution with normal and beta margins respectively is presented. We complete this section with details on maximum likelihood estimation.

\subsection{\label{overview}Overview and relevant background for vine copulas}
A copula is a multivariate cdf with uniform $U(0,1)$ margins. \cite{joe97,joe2014,nelsen06}
If $F$ is a $d$-variate cdf with univariate margins $F_1,\ldots,F_d$,
then Sklar's\cite{sklar1959}  theorem implies that there is a copula $C$ such that
  $$F(x_1,\ldots,x_d)= C\Bigl(F_1(x_1),\ldots,F_d(x_d)\Bigr).$$
The copula is unique if $F_1,\ldots,F_d$ are continuous.
If $F$ is continuous and $(Y_1,\ldots,Y_d)\sim F$, then the unique copula
is the distribution of $(U_1,\ldots,U_d)=\left(F_1(Y_1),\ldots,F_d(Y_d)\right)$ leading to
  $$C(u_1,\ldots,u_d)=F\Bigl(F_1^{-1}(u_1),\ldots,F_d^{-1}(u_d)\Bigr),
  \quad 0\le u_j\le 1, j=1,\ldots,d,$$
where $F_j^{-1}$ are inverse cdfs.\cite{nikoloulopoulos&joe12} In particular,
if $\Phi_d(\cdot;\R)$
is the MVN cdf with correlation matrix $$\R=(\rho_{jk}: 1\le j<k\le d)$$ and
N(0,1) margins, and $\Phi$ is the univariate standard normal cdf,
then the MVN copula is
\begin{equation}\label{MVNcdf}
C(u_1,\ldots,u_d)=\Phi_d\Bigl(\Phi^{-1}(u_1),\ldots,\Phi^{-1}(u_d);\R\Bigr).
\end{equation}

A copula $C$ has reflection symmetry if  $(U_1,\ldots,U_d)\sim C$
implies that $(1-U_1,\ldots,1-U_d)$ has the same distribution $C$. This is apparently the case for the MVN copula. 
When it is necessary to have copula models with reflection asymmetry
and flexible lower/upper tail dependence, then vine copulas
are the best choice.\cite{joeetal10} The
$d$-dimensional vine copulas are built via successive mixing from
$d(d-1)/2$ bivariate linking copulas on trees and their cdfs
involve lower-dimensional integrals. Since the densities of multivariate
vine copulas can be factorized in terms of bivariate linking copulas
and lower-dimensional margins, they are computationally tractable.
Depending on the types of trees, various vine copulas can be
constructed. Two  
types are D-vines and
C-vines.

\begin{figure}[!h]
\vspace{3cm}
\begin{center}
\setlength{\unitlength}{1.cm}
\begin{picture}(5,1)
\put(0,0.5){\framebox(0.5,0.5){$1$}}
\put(2.5,3){\framebox(0.5,0.5){$2$}}
\put(0.5,0.8){\line(1,1){2.2}}
\put(1.3,2.){\makebox(0,0){$12$}}
\put(2.5,0.5){\framebox(0.5,0.5){$3$}}
\put(0.5,0.8){\line(1,0){2}}
\put(1.5,1.){\makebox(0,0){$13$}}
\put(1.5,0){$T_1$}
\put(3.5,0.5){\framebox(0.5,0.5){$12$}}
\put(6.,0.5){\framebox(0.5,0.5){$13$}}
\put(4,0.8){\line(1,0){2}}
\put(5,1){\makebox(0,0){$23|1$}}
\put(4.8,0){$T_2$}

\end{picture}
\end{center}

\caption{\label{3cvine1}3-dimensional C-vine with 2 trees/levels
($T_j$, $j=1,2$) where the dependence is driven by $Y_1$.}
\end{figure}
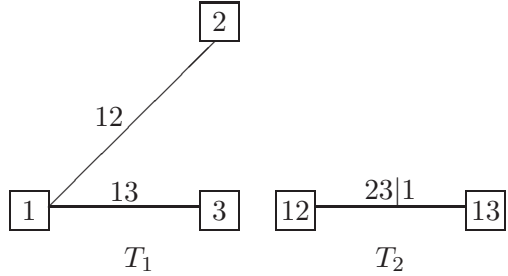

For the $d$-dimensional D-vine, the pairs at level 1 are $i,i+1$, for
$i=1,\ldots,d-1$, and for level $\ell$ ($2\le\ell<d$), the (conditional)
pairs are $i,i+\ell|i+1,\ldots,i+\ell-1$ for $i=1,\ldots,d-\ell$.
For the $d$-dimensional C-vine, the pairs at level 1 are $1,i$, for
$i=2,\ldots,d$, and for level $\ell$ ($2\le\ell<d$), the (conditional)
pairs are $\ell,i|1,\ldots,\ell-1$ for $i=\ell+1,\ldots,d$.
That is, for the D-vine, conditional copulas are specified
for variables $i$ and $i+\ell$ given the variables indexed in between;
and for the C-vine, conditional copulas are specified for variables $\ell$ and
$i$ given those indexed as 1 to $\ell-1$.\cite{nikoloulopoulos&joe&li11}
In Figures  \ref{3dvine} and \ref{3cvine1}, a D-vine and a C-vine with 3 variables and 2 trees/levels are
depicted. From the graphs is apparent that the D- and C-vine structures in the three-dimensional case coincide.
The trivariate density is decomposed in a simple manner by multiplying the nodes of the
nested set of trees
\begin{eqnarray}\label{vine-density}
  f_{123}(x_1,x_2,x_3)&=&f_2(x_2)f_1(x_1)f_3(x_3)\nonumber c_{12}\bigl(F_1(x_1),F_2(x_2)\bigr)\,c_{13}\bigl(F_1(x_1),F_3(x_3)\bigr)c_{23|1}\bigl(F_{2|1}(x_2|x_1),F_{3|1}(x_3|x_1)\bigr)\nonumber\\
&=&f_1(x_1)f_2(x_2)f_3(x_3)c_{123}\bigl(F_1(x_1),F_2(x_2),F_3(x_3)\bigr),
\end{eqnarray}
where $F_{j|k}(x_j|x_k)=\partial C_{jk}\bigl(F_j(x_j),F_k(x_k)\bigr)/\partial F_k(x_k)$. \cite{joe96}

\begin{figure}[!h]
\vspace{1cm}
\begin{center}
\setlength{\unitlength}{1.cm}
\begin{picture}(5,1)
\put(1.2,0.7){\framebox(0.5,0.5){$2$}}
\put(1.7,1){\line(1,0){1}}
\put(2.2,1.2){\makebox(0,0){$12$}}
\put(2.7,0.7){\framebox(0.5,0.5){$1$}}
\put(3.2,1){\line(1,0){1}}
\put(3.7,1.2){\makebox(0,0){$13$}}
\put(4.2,0.7){\framebox(0.5,0.5){$3$}}
\put(5,0.9){$T_1$}
\put(2.,-0.5){\framebox(0.5,0.5){$12$}}
\put(2.5,-0.25){\line(1,0){1}}
\put(3.,-0.){\makebox(0,0){\small$23|1$}}
\put(3.5,-0.5){\framebox(0.5,0.5){$13$}}
\put(5,-0.3){$T_2$}
\end{picture}
\end{center}

\caption{\label{3dvine}3-dimensional D-vine with 2 trees/levels
($T_j$, $j=1,2$).}
\end{figure}
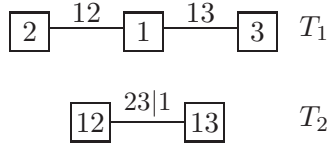

\subsection{\label{normal-parametrization}The copula mixed  model for the latent pair of transformed sensitivity and specificity}
Here we generalize the GLMM by  proposing a model that links the three random effects using a copula function rather than the TVN distribution.

The within-study model is the same as in the standard GLMM; see (\ref{withinBinom}).
The stochastic representation of the between studies model takes the form
\begin{equation}\label{copula-between-norm}
\Bigl(\Phi\bigl(X_1;l(\pi_1),\s_1^2\bigr),\Phi\bigl(X_2;l(\pi_2),\s_2^2\bigr)
,\Phi\bigl(X_3;l(\pi_3),\s_3^2\bigr)\Bigr)\sim C(\cdot;\thbf),
\end{equation}
where $C(\cdot;\thbf)$ is a vine  copula with dependence parameter vector $\thbf=(\th_{12},\th_{13},\th_{23|1})$ and $\Phi(\cdot;\mu,\s^2)$ is the cdf of the  N($\mu,\s^2$) distribution.
The joint density $f_{123}(x_1,x_2,x_3)$ of the transformed latent proportions can be derived using the vine density representation in (\ref{vine-density}):
\begin{multline}\label{jointdensityNCMM}
f_{123}(x_1,x_2,x_3;\pi_1,\pi_2,\pi_3,\s_1,\s_2,\s_3,\thbf)=\\
c_{123}\Bigl(\Phi\bigl(x_1;l(\pi_1),\s_1^2\bigr),\Phi\bigl(x_2;l(\pi_2),\s_2^2\bigr)
,\Phi\bigl(x_3;l(\pi_3),\s_3^2\bigr);\thbf\Bigr)\prod_{j=1}^3\phi\bigl(x_j;l(\pi_j),\s_j^2\bigr),
\end{multline}
where $\phi(\cdot;\mu,\s^2)$ is the  N($\mu,\s^2$)  density.
The models in (\ref{withinBinom}) and (\ref{copula-between-norm}) together specify a vine copula mixed  model with joint likelihood

\begin{eqnarray}\label{mixed-cop-likelihood}
L(\pi_1,\pi_2,\s_1,\s_2,\thbf)&=&\prod_{i=1}^N\int_{-\infty}^{\infty}\int_{-\infty}^{\infty}\int_{-\infty}^{\infty}
\prod_{j=1}^3g\Bigl(y_{ij};n_{ij},l^{-1}(x_j)\Bigr)
c_{123}\Bigl(\Phi\bigl(x_1;l(\pi_1),\s_1^2\bigr),\Phi\bigl(x_2;l(\pi_2),\nonumber\\
&&\qquad\qquad\qquad\qquad \s_2^2\bigr)\Phi\bigl(x_3;l(\pi_3),\s_3^2\bigr);\thbf\Bigr)\prod_{j=1}^3\phi\bigl(x_j;l(\pi_j),\s_j^2\bigr)dx_j\\
&=&
\prod_{i=1}^N\int_{0}^{1}\int_{0}^{1}\int_{0}^{1}
\prod_{j=1}^3g\Bigl(y_{ij};n_{ij},l^{-1}\bigl(\Phi^{-1}(u_j;l(\pi_j),\s_j^2)\bigr)\Bigr)c_{123}(u_1,u_2,u_3;\thbf)du_j.\nonumber
\end{eqnarray}

\subsubsection{Relationship with the GLMM}

In this subsection, we show what happens when  the bivariate copula block is the BVN copula.
The resulting distribution is the TVN with  variance-covariance matrix $\Sbf$, where 
$\rho_{23}=\rho_{23|1}\sqrt{1-\rho_{12}^2}\sqrt{1-\rho_{13}^2}+\rho_{12}\rho_{13}$.\cite{aasetal09} 
In fact, the correlation parameter $\rho_{23|1}$  of  $c_{23|1}$,  is the  partial correlation of $X_2$ and $X_3$ given $X_1$.

\subsection{\label{beta-parametrization}The vine copula mixed  model for the latent pair of sensitivity and specificity}

The within-study model assumes that the number of true positives $Y_{i1}$,  true negatives $Y_{i2}$, and diseased persons  $Y_{i3}$ are conditionally independent and binomially distributed given $\X=\x$, where $\X=(X_1,X_2,X_3)$ denotes the  bivariate latent random pair of sensitivity and specificity.
That is
\begin{eqnarray}\label{withinBinom2}
Y_{i1}|X_{1}=x_1&\sim& \mbox{Binomial}(n_{i1},x_1);\nonumber\\
Y_{i2}|X_{2}=x_2&\sim& \mbox{Binomial}(n_{i2},x_2);\\
Y_{i3}|X_{3}=x_3&\sim& \mbox{Binomial}(n_{i3},x_3).\nonumber
\end{eqnarray}
So one  does not have to transform the latent sensitivity and specificity and can work on the original scale.
The Beta($\a,\b$) distribution can be used for the marginal modeling of
the latent proportions and its density is
$$f(x;\a,\b)=\frac{x^{\a-1}(1-x)^{\b-1}}{B(\a,\b)},\quad 0<x<1, \quad \a,\b>0.$$
In the sequel we will use the
Beta($\pi,\gamma$) parametrization, where $\pi=\frac{\a}{\a+\b}$ (mean parameter) and $\g=\frac{1}{\a+\b+1}$ (dispersion parameter).

The stochastic representation of the between studies model is
\begin{equation}\label{copula-between}
\Bigl(F(X_1;\pi_1,\g_1),F(X_2;\pi_2,\g_2),F(X_3;\pi_3,\g_3)\Bigr)\sim C(\cdot;\thbf),
\end{equation}
where $C(\cdot;\thbf)$ is a  vine copula with dependence parameter vector $\thbf$ and $F(\cdot;\pi,\g)$ is the cdf of the the Beta($\pi,\g$) distribution.
The models in (\ref{withinBinom2}) and (\ref{copula-between}) together specify a vine copula mixed  model with joint likelihood
\begin{eqnarray}\label{beta-mixed-cop-likelihood}
L(\pi_1,\pi_2,\pi_3,\g_1,\g_2,\g_3,\thbf)&=&\prod_{i=1}^N\int_0^1\int_0^1\int_0^1
\prod_{j=1}^3g(y_{ij};n_{ij},x_j)
c_{123}\Bigl(F(x_1;\pi_1,\g_1),F(x_2;\pi_2,\g_2),F(x_3;\nonumber\\
&&\qquad\qquad\qquad\qquad\qquad\qquad\pi_3,\g_3);\thbf\Bigr)
\prod_{j=1}^3f(x_j;\pi_j,\g_j)dx_j\\
&=&\prod_{i=1}^N\int_0^1\int_0^1\int_0^1
\prod_{j=1}^3g\bigl(y_{ij};n_{ij},F^{-1}(u_j;\pi_j,\g_j)\bigr)c(u_1,u_2,u_3;\thbf)du_j.\nonumber
\end{eqnarray}

In the vine copula mixed model with beta (normal) margins, it is important to note that the copula parameter vector $\thbf$ contains parameters of the random effects model and it is separated from the univariate parameters. The univariate parameters $\pi_j$'s  are those of actual interest denoting the meta-analytic parameters for the sensitivity, specificity, and disease prevalence, while the univariate parameters $\sigma_j$'s ($\g_j$'s) are of secondary interest expressing the variability between studies.

\subsubsection{Relationship with existing models}

Hoyer and Kuss  \cite{hoyer&kuss2015}  proposed a vine copula model with beta-binomial margins in this context. This model is actually an approximation of the copula mixed model with beta margins for the latent pair of sensitivity and specificity in (\ref{withinBinom2}) and (\ref{copula-between}).
They attempt to approximate the likelihood  in (\ref{beta-mixed-cop-likelihood})  with the likelihood of a copula model for observed discrete variables which have beta-binomial margins.

The approximation that they suggest is 
\begin{multline*}
L(\pi_1,\pi_2,\pi_3,\g_1,\g_2,\g_3,\thbf)\approx\\
\prod_{i=1}^Nc_{123}\Bigl(H(y_{i1};n_{i1},\pi_1,\g_1),H(y_{i2};n_{i2},\pi_2,\g_2)
,H(y_{i3};n_{i3},\pi_3,\g_3);\thbf\Bigr)\prod_{j=1}^3 h(y_{ij};n_{ij},\pi_j,\g_j),
\end{multline*}
where $H(\cdot;n,\pi,\g)$ is the cdf of the the Beta-Binomial($n,\pi,\g$) distribution.
In their approximation the authors also treat the observed variables which have beta-binomial distributions as being continuous, and model them under the theory for copula models with continuous margins. They state that the computations are becoming too complex in the trivariate case. This is far from true; for $d=3$ the evaluation of the discrete likelihood can be handled extremely easily. 
Copula modelling for $d$-dimensional discrete data  
is studied by Nikoloulopoulos \cite{nikoloulopoulos13b,nikoloulopoulos2015} for MVN copulas and by Panagiotelis {\it et al.} \cite{panagiotelis&czado&joe12} and Nikoloulopoulos and Joe \cite{nikoloulopoulos&joe12}   for vine copulas for a dimension $d$ much higher than 3.

The main problem in Hoyer and Kuss\cite{hoyer&kuss2015} (even if treating the observed variables which have beta-binomial distributions as being discrete) is that 
the discrete  $Y_{ij}$ do not have a common support over different studies or $i$, hence, one cannot conclude that there is a copula for $(Y_{i1},Y_{i2},Y_{i3})$ that applies when the $n_{ij}$ vary with different studies $i$. The natural replicability is in the random effects probability for sensitivity, specificity, and disease prevalence. 
Nikoloulopoulos \cite{nikoloulopoulos2015remada}  has  shown in the bivariate case  that this approximation\cite{kuss-etal-2013} is an  inefficient approach;  it leads  to substantial downward bias for the estimates  of the copula parameters and bias for the meta-analytic parameters for fully specified copula mixture models.  

Therefore, the approximation  method in Hoyer and Kuss\cite{hoyer&kuss2015}   is not discussed/used in the sequel since its inefficiency with theoretical and small-sample efficiency calculations has been proven  \cite{nikoloulopoulos2015remada} and should be avoided for meta-analysis of diagnostic accuracy test studies.

\subsection{\label{computation}Maximum likelihood estimation and computational details}

Estimation of the model parameters $(\pi_1,\pi_2,\pi_3,\s_1,\s_2,\s_3,\thbf)$ and $(\pi_1,\pi_2,\pi_3,\g_1,\g_2,\g_3,\thbf)$   can be approached by the standard maximum likelihood (ML) method, by maximizing the logarithm of the joint likelihood in (\ref{mixed-cop-likelihood})  and (\ref{beta-mixed-cop-likelihood}), respectively. 
The estimated parameters can be obtained by 
using a quasi-Newton \cite{nash90} method applied to the logarithm of the joint likelihood.  
This numerical  method requires only the objective
function, i.e.,  the logarithm of the joint likelihood, while the gradients
are computed numerically and the Hessian matrix of the second
order derivatives is updated in each iteration. The standard errors (SE) of the ML estimates can be also obtained via the gradients and the Hessian computed numerically during the maximization process. Assuming that the usual regularity conditions  for
asymptotic maximum likelihood theory hold for the trivariate model
as well as for its margins we have that ML estimates are
asymptotically normal.\cite{serfling80} Therefore one can build Wald tests to
statistically judge any effect.

For vine copula mixed models of the form with joint likelihood as in (\ref{mixed-cop-likelihood})  and (\ref{beta-mixed-cop-likelihood}), numerical evaluation of the joint pmf is easily done with the following steps:

\begin{enumerate}
\item Calculate Gauss-Legendre\cite{Stroud&Secrest1966}  quadrature points $\{u_q: q=1,\ldots,n_q\}$ 
and weights $\{w_q: q=1,\ldots,n_q\}$ in terms of standard uniform.
\item Convert from independent uniform random variables $\{u_{q_1}: q_1=1,\ldots,n_q\}$,  $\{u_{q_2}: q_2=1,\ldots,n_q\}$, and $\{u_{q_3}: q_3=1,\ldots,n_q\}$ to dependent uniform random variables $\{v_{q_1}=u_{q_1}: q_1=1,\ldots,n_q\}$, $\bigl\{v_{q_2|q_1}=C^{-1}_{12}(u_{q_2}|u_{q_1};\th_{12}): q_1=q_2=1,\ldots,n_q\bigr\}$, and  
$\Bigl\{v_{q_2q_3|q_1}=C^{-1}_{13}\Bigl(C^{-1}_{23|1}(u_{q_3}|u_{q_2};$ $\th_{23|1})|u_{q_1};\th_{13}\Bigr): q_1=q_2=q_3=1,\ldots,n_q\Bigr\}$ that have vine distribution $C(\cdot;\thbf)$. 
The inverse of the conditional distribution $C(v|u;\th)=\partial C(u,v;\th)/\partial u$ corresponding to the copula $C(\cdot;\th)$ and the simulation algorithm of a C-vine copula in Joe\cite{joe2010a} are used  to achieve this.
\item Numerically evaluate the joint pmf, e.g., 
$$\int_0^1\int_0^1
\prod_{j=1}^3g\bigl(y_{j};n_{j},F^{-1}(u_j;\pi_j,\g_j)\bigr)c(u_1,u_2,u_3;\thbf)du_1du_2du_3$$
in a triple sum:
\end{enumerate}
$$\sum_{q_1=1}^{n_q}\sum_{q_2=1}^{n_q}\sum_{q_3=1}^{n_q}w_{q_1}w_{q_2}w_{q_3}
g\bigl(y_{1};n,F^{-1}(v_{q_1};\pi_1,\g_1)\bigr)g\bigl(y_{2};n,F^{-1}(v_{q_2|q_1};\pi_2,\g_2)\bigr)g\bigl(y_{3};n,F^{-1}(v_{q_2q_3|q_1};\pi_3,\g_3)\bigr).
$$

With Gauss-Legendre quadrature, the same nodes and weights
are used for different functions;
this helps in yielding smooth numerical derivatives for numerical optimization via quasi-Newton.\cite{nash90} 
Our comparisons show that $n_q=15$ is adequate with good precision to at least at four decimal places; thus it also provides the advantage of fast implementation.  To this end, the vine copula mixed  effect model for meta-analysis of diagnostic test accuracy studies  with a triple integral
is in fact straightforward computationally.

\section{Construction of vine copulas} \label{construction}

A  vine copula can be decomposed into a sequence of
bivariate conditional copulas in many ways if the conditional
copulas depend on the values of the conditioning
variables.
Vine copulas involve a sequence of conditional bivariate copulas,
but assume the bivariate conditional copulas are constant over the
conditioning variables. For example, MVN copulas
satisfy this property with conditional correlations being equal to
partial correlations. In general, a given multivariate copula might
be well-approximated by several different vine copulas, and likelihood
inference could be used to find such vines.\cite{nikoloulopoulos&joe&li11}

Vine copulas
can provide a wide range of flexible (tail) dependence  by choosing suitable choices of bivariate
linking copulas,\cite{joeetal10} but they require a decision on the indexing of variables since for a 3-dimensional C-vine or D-vine copula there are $3!/2$ possible different choices (distinct permutations).

In this section, we discuss
the following practical issues for trivariate vine-copula modelling:
\begin{enumerate}
\itemsep=0pt
\item type of bivariate copula families as building blocks;
\item the matching of variables to
labels/indexes;
\item conditional independence. 
\end{enumerate}
These matters have been overseen in Hoyer and Kuss\cite{hoyer&kuss2015}.

 \subsection{\label{blocks}Building blocks}
For fitting trivariate data with possible tail dependence
(in which case, TVN copulas are inappropriate),
we use vine copulas with
parametric families for the bivariate building blocks that have
different strengths of tail behaviour. 
  \cite{nikoloulopoulos2015remada} 
\begin{itemize}
\itemsep=-5pt
\item 
Reflection symmetric copulas with tail independence satisfying
$C(u,u)=O(u^2)$ and $\Cbar(1-u,1-u)=O(u^2)$ as $u\to 0$,
such as the Frank copula with inverse conditional cdf

$$C^{-1}(v|u;\th)=
-\frac{1}{\theta}\log\left[\frac{1-(1-e^{-\th})}{(v^{-1}-1)e^{-\th u}+1}\right]
,\quad \theta \in (-\infty,\infty)\setminus\{0\}.$$
\item 
Reflection symmetric copulas with intermediate tail dependence 
\cite{Hua-joe-11} such as the BVN copula, which satisfies
$C(u,u,\th)=O(u^{2/(1+ \th)}(-\log u)^{-\th/(1+\th)})$ as
$u\to 0$ with inverse conditional cdf
$$C^{-1}(v|u;\th)=\Phi\Bigl(\sqrt{1-\th^2}\Phi^{-1}(v)+\th\Phi^{-1}(u)\Bigr),\quad \theta \in [-1,1].$$
\item
Reflection asymmetric copulas with lower tail dependence satisfying  $c(u,u)=O(u^{-1})$ as $u\to 0$, such as the Clayton copula with inverse conditional cdf
$$C^{-1}(v|u;\th)=
\Bigl\{(v^{-\theta/(1+\theta)}-1)u^{-\th}+1\Bigr\}^{-1/\theta}
,\quad \th\in (0,\infty).$$

\item
Reflection asymmetric copulas with  upper tail dependence satisfying $c(1-u,1-u)=O(u^{-1})$ as $u\to 0$,  such as the
rotated by 180 degrees Clayton copula with
inverse conditional cdf 
$$C^{-1}(v|u;\th)=
1-\Bigl[\bigl\{(1-v)^{-\theta/(1+\theta)}-1\bigr\}(1-u)^{-\th}+1\Bigr]^{-1/\theta}
,\quad \th\in (0,\infty).$$
\item Reflection asymmetric copula family with negative upper-lower tail dependence satisfying $c(1-u,u)=O(u^{-1})$ as $u\to 0$,
such as the rotated by 90 degrees Clayton copula with inverse conditional cdf
$$C^{-1}(v|u;\th)=
\Bigl\{(v^{\theta/(1-\theta)}-1)(1-u)^{\th}+1\Bigr\}^{1/\theta}
,\quad \th\in (0,\infty).$$

\item Reflection asymmetric copula family with negative lower-upper tail dependence satisfying $c(u,1-u)=O(u^{-1})$ as $u\to 0$,
such as the as the rotated by 270 degrees Clayton copula with inverse conditional cdf
$$C^{-1}(v|u;\th)=1-
\Bigl[\bigl\{(1-v)^{\theta/(1-\theta)}-1\bigr\}u^{\th}+1\Bigr]^{1/\theta}
,\quad \th\in (0,\infty).$$

\end{itemize}

The above copula families are sufficient for the applications in Section \ref{sec-appl}, since tail dependence is a property to consider when choosing amongst different families of copulas and the concept of upper/lower tail dependence is one way to differentiate families. Nikoloulopoulos and Karlis \cite{Nikoloulopoulos&karlis08CSDA}
have shown that it is hard to choose a copula with similar properties from real data, since
copulas with similar (tail) dependence properties provide similar fit.
Hoyer and Kuss\cite{hoyer&kuss2015} used, also,  the Placket copula; this bivariate block is not used here since we have included another choice of copulas with similar properties i.e., the Frank parametric family of bivariate copulas.

\subsection{Indexing of the variables}
\begin{figure}[!h]
\vspace{3cm}
\begin{center}
\setlength{\unitlength}{1.cm}
\begin{picture}(5,1)
\put(0,0.5){\framebox(0.5,0.5){$2$}}
\put(2.5,3){\framebox(0.5,0.5){$1$}}
\put(0.5,0.8){\line(1,1){2.2}}
\put(1.3,2.){\makebox(0,0){$12$}}
\put(2.5,0.5){\framebox(0.5,0.5){$3$}}
\put(0.5,0.8){\line(1,0){2}}
\put(1.5,1.){\makebox(0,0){$23$}}
\put(1.5,0){$T_1$}
\put(3.5,0.5){\framebox(0.5,0.5){$12$}}
\put(6.,0.5){\framebox(0.5,0.5){$23$}}
\put(4,0.8){\line(1,0){2}}
\put(5,1){\makebox(0,0){$13|2$}}
\put(4.8,0){$T_2$}
\end{picture}
\end{center}

\caption{\label{3cvine2}3-dimensional C-vine with 2 trees/levels
($T_j$, $j=1,2$) where the dependence is driven by $Y_2$.}
\end{figure}
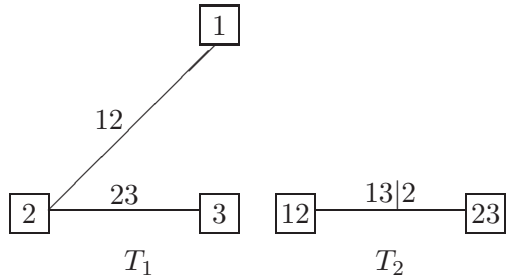

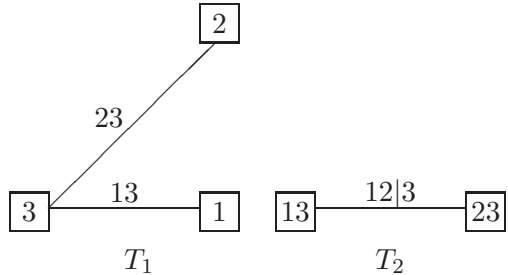
\begin{figure}[!h]
\vspace{2cm}
\begin{center}
\setlength{\unitlength}{1.cm}
\begin{picture}(5,1)
\put(0,0.5){\framebox(0.5,0.5){$3$}}
\put(2.5,3){\framebox(0.5,0.5){$2$}}
\put(0.5,0.8){\line(1,1){2.2}}
\put(1.3,2.){\makebox(0,0){$23$}}
\put(2.5,0.5){\framebox(0.5,0.5){$1$}}
\put(0.5,0.8){\line(1,0){2}}
\put(1.5,1.){\makebox(0,0){$13$}}
\put(1.5,0){$T_1$}
\put(3.5,0.5){\framebox(0.5,0.5){$13$}}
\put(6.,0.5){\framebox(0.5,0.5){$23$}}
\put(4,0.8){\line(1,0){2}}
\put(5,1){\makebox(0,0){$12|3$}}
\put(4.8,0){$T_2$}
\end{picture}
\end{center}

\caption{\label{3cvine3}3-dimensional C-vine with 2 trees/levels
($T_j$, $j=1,2$) where the dependence is driven by $Y_3$.}
\end{figure}

The next step would be to decide on the bivariate pairs
to put at level 1 of the vine.
Given the small dimension ($d=3$) we can 
fit all possible permutations of the vine  and select the best based on the likelihood principle. 
In Figures  \ref{3cvine1}, \ref{3cvine2},    and \ref{3cvine3} we depict the three distinct permutations: $$\{12,13,23|1\}, \qquad \{12,23,13|2\}, \quad \mbox{and} \quad \{13,23,12|3\}$$ using  a C-vine structure. 

\subsection{\label{condind}Conditional independence}
Joe {\it et al.}\cite{joeetal10} show that in order for a vine copula to have (tail) dependence for all bivariate margins, it is only necessary for the bivariate copulas in level 1 to have (tail) dependence and it is not necessary for the conditional bivariate copulas in levels $2,\ldots,d-1$ to have tail dependence. That provides the theoretical justification for the idea to capture the strongest dependence in the first tree and then just use the independence copulas in lower order tree(s), i.e., conditional independence. 

The improvement over the reduction of the dependence parameters, in the   context of the trivariate meta-analysis of diagnostic test accuracy studies accounting for disease prevalence, is  small (one dependence parameter less), but for estimation purposes this is extremely useful given the typically small sample size.

\section{\label{miss-section}Small-sample efficiency -- Misspecification of the random effects distribution}

An extensive simulation study is conducted  
(a) to gauge the small-sample efficiency of the ML 
method, and 
(b) to investigate in detail 
the  misspecification of the parametric margin or  family of copulas of the random effects distribution.

To simulate the data we have used   the following generation  process:\cite{paul-etal-2010,nikoloulopoulos2015remada}
\begin{enumerate}

\item Simulate the study size $n$ from a shifted gamma distribution, i.e., $n\sim \mbox{sGamma}(\a=1.2,\b=0.01,\mbox{lag}=30)$ and round off to the nearest integer. 
\item Simulate $(u_1,u_2,u_3)$ from a C-vine $C(;\tau_{12},\tau_{13},\tau_{23|1})$ via the algorithm  in Joe \cite{joe2010a};  $\tau$'s are converted 
to the BVN\cite{HultLindskog02}, Frank\cite{genest87} and   rotated  Clayton\cite{genest&mackay86}  $\theta$'s via the relations 
\begin{equation}\label{tauBVN}
\tau=\frac{2}{\pi}\arcsin(\th),
\end{equation}

\begin{equation}\label{tauFrank}
\tau=\left\{\begin{array}{ccc}
1-4\theta^{-1}-4\theta^{-2}\int_\theta^0\frac{t}{e^t-1}dt &,& \th<0\\
1-4\theta^{-1}+4\theta^{-2}\int^\theta_0\frac{t}{e^t-1}dt &,& \th>0\\
\end{array}\right.,
\end{equation}

\begin{equation}\label{tauCln}
\tau=\left\{\begin{array}{rcl}
\th/(\th+2)&,& \mbox{by 0 or 180 degrees}\\
-\th/(\th+2)&,& \mbox{by 90 or 270 degrees}\\
\end{array}\right.,
\end{equation}
respectively.

\item Convert to beta realizations  via $x_j=F_j^{-1}(u_j,\pi_j,\g_j)$ or normal realizations via $x_j=\Phi_j^{-1}\bigl(u_j,l(\pi_j),\s_j\bigr)$ for $j=1,2$; for the latter convert to proportions via $x_j=l^{-1}(x_j)$.    
\item Set number of diseased and non-diseased as $n_{1}=nx_1$ and $n_2=n-n_1$, respectively. 
\item Set   $y_j=n_jx_j$ and then round $y_j$ for $j=1,2$. 
\end{enumerate}

We randomly generated $B=10^4$ samples of size  $N = 20$ from  the C-vine  copula mixed model with Clayton blocks rotated by 90 degrees,  beta margins, and conditional independence.
Table \ref{sim}  contains the 
resultant biases, root mean square errors (RMSE), and standard deviations (SD), along with average theoretical SDs $(\sqrt{\bar V})$ scaled by $N$, for the MLEs  under different block choices and margins assuming conditional  at the second level/tree, i.e., we fix $\tau_{23|1}=0$. The theoretical variances $V$ of the MLEs are obtained via the gradients and the Hessian computed
numerically during the maximization process.

Conclusions from the values in the table and ohter computations that we have done are the following:
\begin{itemize}

\item ML  with  the the `true' vine copula mixed  model is highly efficient according to the simulated biases and standard deviations for the meta-analytic parameters of interest.

\end{itemize}

\begin{landscape}
\begin{table}[htbp]
  \centering
  \caption{\label{sim} Small sample of sizes $N = 20$ simulations ($10^4$ replications) from the vine copula mixed model with Clayton rotated by 90 degrees blocks, beta margins, and conditional independence, resultant biases,  root mean square errors (RMSE) and standard deviations (SD), along with the square root of the average theoretical variances ($\sqrt{\bar V}$), scaled by $100$, for the MLEs  under different copula choices and margins. Vines have been truncated at the 1st level/tree, i.e. $\tau_{23|1}$ is fixed to 0 (conditional independence).}
    \begin{tabular}{ccccccccccc}
    \hline
          & Margin & Block & $\pi_1=0.8$ & $\pi_2=0.7$ & $\pi_3=0.4$ & $\g_1=0.1$ & $\g_2=0.1$ & $\g_3=0.05$ & $\tau_{12}=-0.5$ & $\tau_{13}=-0.3$ \\
    \hline
    Bias  & Beta  & BVN   & 0.04  & 0.00  & 0.02  & -2.28 & -2.01 & -1.11 & -12.24 & -6.17 \\
          &       & Frank & 0.97  & -1.12 & -0.59 & -2.48 & -1.85 & -1.09 & -14.81 & -7.15 \\
          &       & Clayton by 90 & 0.07  & 0.06  & 0.05  & -2.52 & -2.14 & -1.18 & -14.66 & -6.35 \\
          &       & Clayton by 270 & 0.19  & -0.56 & -0.14 & -1.91 & -1.38 & -0.90 & -8.85 & 2.36 \\
          & Normal & BVN   & 2.43  & 1.77  & -0.39 & - & - & - & -12.16 & -6.75 \\
          &       & Frank & 3.19  & 0.52  & -1.00 & - & - & -& -14.57 & -7.45 \\
          &       & Clayton by 90 & 2.39  & 1.64  & -0.44 & - & - & - & -14.82 & -6.73 \\
          &       & Clayton by 270 & 2.57  & 1.38  & -0.52 & - & - & - & -8.97 & 1.15 \\\hline
    SD    & Beta  & BVN   & 2.95  & 3.29  & 2.50  & 3.22  & 2.74  & 1.47  & 18.81 & 19.59 \\
          &       & Frank & 2.99  & 3.70  & 2.69  & 3.11  & 2.86  & 1.50  & 18.82 & 20.53 \\
          &       & Clayton by 90 & 2.93  & 3.31  & 2.53  & 2.89  & 2.62  & 1.42  & 17.31 & 19.51 \\
          &       & Clayton by 270 & 3.07  & 3.58  & 2.61  & 3.85  & 3.33  & 1.68  & 27.99 & 24.01 \\
          & Normal & BVN   & 3.01  & 3.49  & 2.61  & 18.67 & 13.83 & 8.74  & 18.88 & 19.67 \\
          &       & Frank & 3.16  & 3.84  & 2.80  & 18.78 & 14.10 & 8.87  & 18.86 & 20.55 \\
          &       & Clayton by 90 & 3.01  & 3.53  & 2.64  & 17.46 & 13.11 & 8.61  & 17.51 & 19.84 \\
          &       & Clayton by 270 & 3.12  & 3.67  & 2.70  & 21.58 & 16.44 & 9.69  & 27.80 & 23.38 \\\hline
    $\sqrt{\bar V}$ & Beta  & BVN   & 2.65  & 3.06  & 2.36  & 2.78  & 2.59  & 1.43  & 17.59 & 15.91 \\
          &       & Frank & 2.63  & 3.22  & 2.35  & 2.42  & 2.68  & 1.43  & 5.21  & 16.60 \\
          &       & Clayton by 90 & 2.60  & 2.94  & 2.30  & 2.60  & 2.37  & 1.34  & 13.86 & 16.59 \\
          &       & Clayton by 270 & 2.61  & 3.11  & 2.38  & 2.90  & 2.84  & 1.53  & 14.12 & 14.28 \\
          & Normal & BVN   & 2.73  & 3.12  & 2.42  & 15.88 & 11.86 & 7.83  & 16.73 & 17.34 \\
          &       & Frank & 2.66  & 3.30  & 2.42  & 15.97 & 12.75 & 7.95  & 14.93 & 16.76 \\
          &       & Clayton by 90 & 2.64  & 3.09  & 2.37  & 15.27 & 11.32 & 7.60  & 11.74 & 16.01 \\
          &       & Clayton by 270 & 2.76  & 3.26  & 2.49  & 16.11 & 13.30 & 8.32  & 15.77 & 17.11 \\\hline
    RMSE  & Beta  & BVN   & 2.95  & 3.29  & 2.50  & 3.94  & 3.40  & 1.84  & 22.44 & 20.54 \\
          &       & Frank & 3.14  & 3.87  & 2.76  & 3.97  & 3.40  & 1.85  & 23.95 & 21.74 \\
          &       & Clayton by 90 & 2.94  & 3.31  & 2.53  & 3.83  & 3.39  & 1.84  & 22.68 & 20.52 \\
          &       & Clayton by 270 & 3.08  & 3.63  & 2.61  & 4.30  & 3.60  & 1.91  & 29.36 & 24.12 \\
          & Normal & BVN   & 3.87  & 3.91  & 2.64  & - & - & - & 22.46 & 20.80 \\
          &       & Frank & 4.49  & 3.88  & 2.97  & - & - & -& 23.84 & 21.86 \\
          &       & Clayton by 90 & 3.84  & 3.89  & 2.68  & - & - & -& 22.95 & 20.96 \\
          &       & Clayton by 270 & 4.04  & 3.92  & 2.74  & - & - & - & 29.21 & 23.41 \\
    \hline
    \end{tabular}%
  \label{tab:addlabel}%
\end{table}
\end{landscape}

\begin{itemize}

\item The MLEs of the meta-analytic parameters are slightly underestimated under copula misspecification. That is, there is some downward bias for these parameters, especially if the ``working" model is not close to Kullback-Liebler distance with the ``true" model, i.e., it is misspecified. For example in the table there is more bias for the Clayton rotated by 270 degrees and Frank blocks since they have different tail dependence from the `true' model, i.e., the rotated Clayton by 270 degrees. An interesting result is that the BVN copula performed rather well under misspecification.

\item The SDs are rather robust to the copula misspecification.

\item The meta-analytic MLEs and SDs are not robust to the margin misspecification, while 
the MLE of $\tau$ and  its SD is.

\item Due to small sample size ($N=20$) there is about  2\% and 10\%  bias for the  variability and dependence parameters, respectively. These biases decrease as $N$ increases.   

\end{itemize}

The effect of misspecifying the copula choice can be seen  as minimal for both the univariate parameters and Kendall's tau. However, note that (a) the meta-analytic parameters are a univariate inference, and hence it is the univariate marginal distribution that  matters and not the type of the copula, and, (b) as previously emphasized
Kendall's tau only accounts for the dependence dominated by the middle of the data (sensitivities and specificities), and it is expected to be similar amongst  different families of copulas. However, the tail dependence varies, as explained in Subsection \ref{blocks}, and is a property to consider when choosing amongst different families of copulas, and, hence affects the shape of SROC curves, i.e., prediction.  SROC  will essentially show the
effect of different  model (random effect distribution) assumptions, since it is an inference that depends on the joint distribution. We refer the interested reader to Nikoloulopoulos \cite{nikoloulopoulos2015remada} for further details.

\section{\label{sec-appl}Illustrations}
We illustrate the use of the vine copula mixed model for the meta-analysis of diagnostic accuracy studies accounting from disease prevalence by re-analysing
the data of two published meta-analyses.\cite{Karageorgopoulos-etal-2011,ye-etal-2012} These data have been  previously analyzed in the trivariate case by  Hoyer and Kuss. \cite{hoyer&kuss2015}

We fit the vine copula mixed model for all different permutations, choices of parametric families of copulas and margins.
To make it easier
to compare strengths of dependence, we convert from $\tau$ to the BVN\cite{HultLindskog02}, Frank\cite{genest87} and   rotated  Clayton\cite{genest&mackay86} copula parameter $\theta$ via the relations in (\ref{tauBVN}), (\ref{tauFrank}), and (\ref{tauCln}).

Model selection is often based on information criteria such as AIC and BIC.  We adopt one of this criteria, namely the AIC. The discussion below
could also apply to other information criteria. 
By using the ML method,
the AIC is $-2\times$log-likelihood $+2\times$ (\#model parameters)
and a smaller AIC value indicates a better fitting model.  

We further compute the Vuong's  test\cite{vuong1989} to reveal if any other vine copula mixed model provides better fit than the standard trivariate GLMM as in Nikoloulopoulos.\cite{nikoloulopoulos2015remada}
The Vuong's test is the sample version of the difference in Kullback-Leibler divergence between two models and can be used to differentiate two  parametric models which could be non-nested. 
Assume that we have Models 1 and 2 with parametric densities $f^{(1)}$ and  $f^{(2)}$ respectively,  with Model 1 being the vine copula mixed model composed by BVN copulas and normal margins, i.e., the standard GLMM. 
The sample version  of the difference in Kullback-Leibler divergence between two models with MLEs $\hat\thbf^{(1)},\hat\thbf^{(2)}$ is
$$\bar D=\sum_{i=1}^N D_i/N,$$
where $D_i=\log\left[\frac{f^{(2)}\left(Y_1,Y_2;\hat\thbf^{(2)}\right)}{f^{(1)}\left(Y_1,Y_2;\hat\thbf^{(1)}\right)}\right]$.
Model 1  is the better fitting model if $\bar D<0$, and Model 2 is the better fitting model if $\bar D>0$. 
Vuong \cite{vuong1989} 
has shown that asymptotically under the null hypothesis that  Models 1 and 2 have the same  parametric densities $f^{(1)}$ and  $f^{(2)}$,  
$$z_0=\sqrt{N}\bar D/s\widesim{H_0}\mathcal{N}(0,1),$$
where  $s^2=\frac{1}{N-1}\sum_{i=1}^N(D_i-\bar D)^2$.

\subsection{The $\beta$-D-Glucan data}

In this section we apply the copula mixed models  to data on the 8 cohort studies  in the meta-analysis in Karageorgopoulos {\it et al.}\cite{Karageorgopoulos-etal-2011}. The interest here is to assess $\beta$-D-Glucan  as a serum or plasma marker for the presence of invasive fungal infections, where different medical criteria are used as reference standards.

The AICs showed that a vine copula mixed model with rotated Clayton copulas by 0 (for positive dependence) and 90 (for negative dependence) degrees, beta margins, and permutation $\{12,23,13|2\}$ provides the best fit (Table \ref{perm2-glucan}). It is revealed that a vine copula mixed model with the sensitivity, specificity, and disease prevalence on the original scale provides better fit than the GLMM, which models the sensitivity, specificity and disease prevalence on a transformed scale. In fact,  from the Vuong's statistic there is enough improvement to get a statistical significant difference ($p$-value$= 0.008$).

Due to small sample size in this study 
we have then fitted 
all the truncated at level-1 vines copula mixed models with different permutations and choices of parametric families of copulas and margins.

\begin{landscape}
\begin{table}[htbp]
  \centering
  \caption{\label{perm2-glucan}AICs, estimates and standard errors (SE), along with the Vuong's statistics and $p$-values for the $\beta$-D-Glucan data where the dependence is driven by the number of true negatives $Y_2$.}
    \begin{tabular}{ccccccccccccc}
    \hline
    \multicolumn{13}{c}{Normal margins} \\
    \hline
          & \multicolumn{2}{c}{BVN} & \multicolumn{2}{c}{Frank} & \multicolumn{2}{c}{Clayton by 90/0} & \multicolumn{2}{c}{Clayton by 270/0} & \multicolumn{2}{c}{Clayton by 90/180} & \multicolumn{2}{c}{Clayton by 270/180} \\
          & Est.  & SE    & Est.  & SE    & Est.  & SE    & Est.  & SE    & Est.  & SE    & Est.  & SE \\\hline
    $\pi_2$ & 0.87  & 0.04  & 0.88  & 0.03  & 0.87  & 0.03  & 0.87  & 0.03  & 0.87  & 0.03  & 0.87  & 0.03 \\
    $\pi_3$ & 0.14  & 0.03  & 0.13  & 0.02  & 0.14  & 0.02  & 0.14  & 0.03  & 0.14  & 0.02  & 0.14  & 0.03 \\
    $\pi_1$ & 0.71  & 0.06  & 0.71  & 0.06  & 0.70  & 0.06  & 0.71  & 0.07  & 0.71  & 0.06  & 0.71  & 0.07 \\
    $\s_2$ & 0.69  & 0.21  & 0.70  & 0.21  & 0.68  & 0.21  & 0.70  & 0.21  & 0.68  & 0.21  & 0.70  & 0.21 \\
    $\s_3$ & 0.52  & 0.16  & 0.53  & 0.16  & 0.49  & 0.15  & 0.57  & 0.18  & 0.49  & 0.14  & 0.56  & 0.18 \\
    $\s_1$ & 0.71  & 0.26  & 0.71  & 0.26  & 0.70  & 0.26  & 0.75  & 0.31  & 0.72  & 0.28  & 0.75  & 0.30 \\
    $\tau_{23}$ & -0.41 & 0.23  & -0.37 & 0.22  & -0.35 & 0.25  & -0.34 & 0.22  & -0.39 & 0.23  & -0.37 & 0.22 \\
    $\tau_{12}$ & 0.08  & 0.31  & 0.09  & 0.29  & 0.04  & 0.23  & 0.17  & 0.23  & 0.00  & 0.26  & 0.16  & 0.25 \\
    $\tau_{13|2}$ & -0.12 & 0.31  & -0.10 & 0.29  & -0.22 & 0.25  & -0.30 & 0.22  & -0.10 & 0.23  & -0.02 & 0.25 \\\hline
    AIC   & \multicolumn{2}{c}{188.06} & \multicolumn{2}{c}{188.49} & \multicolumn{2}{c}{187.23} & \multicolumn{2}{c}{188.00} & \multicolumn{2}{c}{187.51} & \multicolumn{2}{c}{188.75} \\\hline
    \multicolumn{13}{c}{Vuong's test} \\\hline
    $z_0$ & \multicolumn{2}{c}{-} & \multicolumn{2}{c}{-0.81} & \multicolumn{2}{c}{1.45} & \multicolumn{2}{c}{-1.08} & \multicolumn{2}{c}{0.99} & \multicolumn{2}{c}{-1.55} \\
    $p$-value & \multicolumn{2}{c}{-} & \multicolumn{2}{c}{0.42} & \multicolumn{2}{c}{0.15} & \multicolumn{2}{c}{0.28} & \multicolumn{2}{c}{0.32} & \multicolumn{2}{c}{0.12} \\\hline
    \multicolumn{13}{c}{Beta margins} \\\hline
          & \multicolumn{2}{c}{BVN} & \multicolumn{2}{c}{Frank} & \multicolumn{2}{c}{Clayton by 90/0} & \multicolumn{2}{c}{Clayton by 270/0} & \multicolumn{2}{c}{Clayton by 90/180} & \multicolumn{2}{c}{Clayton by 270/180} \\
          & Est.  & SE    & Est.  & SE    & Est.  & SE    & Est.  & SE    & Est.  & SE    & Est.  & SE \\\hline
    $\pi_2$ & 0.85  & 0.03  & 0.86  & 0.03  & 0.86  & 0.03  & 0.85  & 0.03  & 0.86  & 0.03  & 0.85  & 0.03 \\
    $\pi_3$ & 0.15  & 0.02  & 0.14  & 0.02  & 0.15  & 0.02  & 0.15  & 0.03  & 0.15  & 0.02  & 0.15  & 0.03 \\
    $\pi_1$ & 0.69  & 0.06  & 0.69  & 0.06  & 0.68  & 0.06  & 0.69  & 0.06  & 0.69  & 0.06  & 0.69  & 0.06 \\
    $\s_2$ & 0.05  & 0.03  & 0.05  & 0.03  & 0.05  & 0.03  & 0.05  & 0.03  & 0.05  & 0.03  & 0.05  & 0.03 \\
    $\s_3$ & 0.03  & 0.02  & 0.03  & 0.02  & 0.03  & 0.02  & 0.04  & 0.02  & 0.03  & 0.02  & 0.03  & 0.02 \\
    $\s_1$ & 0.09  & 0.06  & 0.09  & 0.06  & 0.09  & 0.05  & 0.10  & 0.07  & 0.09  & 0.06  & 0.10  & 0.06 \\
    $\tau_{23}$ & -0.40 & 0.23  & -0.35 & 0.22  & -0.35 & 0.25  & -0.33 & 0.24  & -0.39 & 0.23  & -0.35 & 0.24 \\
    $\tau_{12}$ & 0.08  & 0.31  & 0.09  & 0.29  & 0.05  & 0.25  & 0.19  & 0.22  & -0.01 & 0.28  & 0.19  & 0.24 \\
    $\tau_{13|2}$ & -0.12 & 0.31  & -0.10 & 0.29  & -0.22 & 0.24  & -0.30 & 0.22  & -0.11 & 0.23  & -0.01 & 0.26 \\\hline
    AIC   & \multicolumn{2}{c}{187.37} & \multicolumn{2}{c}{187.80} & \multicolumn{2}{c}{186.50} & \multicolumn{2}{c}{187.20} & \multicolumn{2}{c}{186.74} & \multicolumn{2}{c}{187.97} \\\hline
    \multicolumn{13}{c}{Vuong's test} \\\hline
    $z_0$ & \multicolumn{2}{c}{0.80} & \multicolumn{2}{c}{0.49} & \multicolumn{2}{c}{2.05} & \multicolumn{2}{c}{-0.63} & \multicolumn{2}{c}{1.48} & \multicolumn{2}{c}{-0.53} \\
    $p$-value & \multicolumn{2}{c}{0.42} & \multicolumn{2}{c}{0.62} & \multicolumn{2}{c}{0.04} & \multicolumn{2}{c}{0.53} & \multicolumn{2}{c}{0.14} & \multicolumn{2}{c}{0.60} \\
    \hline
    \end{tabular}%
  \label{tab:addlabel}%
\end{table}
\end{landscape}

\begin{landscape}
\begin{table}[htbp]
  \centering
  \caption{\label{pars-perm3-glucan}AICs, estimates and standard errors (SE), along with the Vuong's statistics and $p$-values for the $\beta$-D-Glucan data where the dependence is driven by the number of diseased persons $Y_3$. Vines have been truncated at the 1st level/tree, i.e. $\tau_{12|3}$ is fixed to 0 (conditional independence). }
    \begin{tabular}{ccccccccccccc}
    \hline
    \multicolumn{13}{c}{Normal margins} \\
    \hline
          & \multicolumn{2}{c}{BVN} & \multicolumn{2}{c}{Frank} & \multicolumn{2}{c}{Clayton by 90} & \multicolumn{2}{c}{Clayton by 270} & \multicolumn{2}{c}{Clayton by 90/270} & \multicolumn{2}{c}{Clayton by 270/90} \\
          & Est.  & SE    & Est.  & SE    & Est.  & SE    & Est.  & SE    & Est.  & SE    & Est.  & SE \\\hline
    $\pi_3$ & 0.14  & 0.02  & 0.13  & 0.02  & 0.14  & 0.03  & 0.14  & 0.02  & 0.14  & 0.02  & 0.14  & 0.03 \\
    $\pi_1$ & 0.71  & 0.06  & 0.70  & 0.06  & 0.70  & 0.06  & 0.70  & 0.06  & 0.71  & 0.06  & 0.71  & 0.06 \\
    $\pi_2$ & 0.87  & 0.03  & 0.88  & 0.03  & 0.87  & 0.03  & 0.87  & 0.02  & 0.87  & 0.02  & 0.87  & 0.03 \\
    $\s_3$ & 0.52  & 0.16  & 0.53  & 0.17  & 0.58  & 0.19  & 0.49  & 0.14  & 0.50  & 0.15  & 0.56  & 0.18 \\
    $\s_1$ & 0.71  & 0.26  & 0.71  & 0.26  & 0.71  & 0.26  & 0.71  & 0.27  & 0.69  & 0.25  & 0.72  & 0.27 \\
    $\s_2$ & 0.70  & 0.22  & 0.70  & 0.22  & 0.72  & 0.23  & 0.68  & 0.22  & 0.69  & 0.22  & 0.71  & 0.23 \\
    $\tau_{13}$ & -0.14 & 0.30  & -0.11 & 0.28  & -0.26 & 0.28  & -0.05 & 0.25  & -0.21 & 0.27  & -0.09 & 0.30 \\
    $\tau_{23}$ & -0.42 & 0.25  & -0.36 & 0.19  & -0.39 & 0.30  & -0.39 & 0.19  & -0.39 & 0.19  & -0.39 & 0.30 \\\hline
    AIC   & \multicolumn{2}{c}{186.08} & \multicolumn{2}{c}{186.57} & \multicolumn{2}{c}{186.49} & \multicolumn{2}{c}{185.64} & \multicolumn{2}{c}{185.10} & \multicolumn{2}{c}{187.04} \\\hline
    \multicolumn{13}{c}{Vuong's test} \\\hline
   $z_0$  & \multicolumn{2}{c}{28.15} & \multicolumn{2}{c}{2.68} & \multicolumn{2}{c}{1.47} & \multicolumn{2}{c}{2.49} & \multicolumn{2}{c}{4.87} & \multicolumn{2}{c}{2.15} \\
    $p$-value & \multicolumn{2}{c}{$<0.001$} & \multicolumn{2}{c}{0.01} & \multicolumn{2}{c}{0.14} & \multicolumn{2}{c}{0.01} & \multicolumn{2}{c}{$<0.001$} & \multicolumn{2}{c}{0.03} \\\hline
    \multicolumn{13}{c}{Beta margins} \\\hline
          & \multicolumn{2}{c}{BVN} & \multicolumn{2}{c}{Frank} & \multicolumn{2}{c}{Clayton by 90} & \multicolumn{2}{c}{Clayton by 270} & \multicolumn{2}{c}{Clayton by 90/270} & \multicolumn{2}{c}{Clayton by 270/90} \\
          & Est.  & SE    & Est.  & SE    & Est.  & SE    & Est.  & SE    & Est.  & SE    & Est.  & SE \\\hline
    $\pi_3$ & 0.15  & 0.02  & 0.14  & 0.02  & 0.15  & 0.03  & 0.15  & 0.02  & 0.15  & 0.02  & 0.15  & 0.03 \\
    $\pi_1$ & 0.69  & 0.06  & 0.69  & 0.06  & 0.68  & 0.06  & 0.69  & 0.06  & 0.69  & 0.06  & 0.69  & 0.06 \\
    $\pi_2$ & 0.85  & 0.03  & 0.86  & 0.03  & 0.85  & 0.03  & 0.86  & 0.03  & 0.86  & 0.03  & 0.86  & 0.03 \\
    $\g_3$ & 0.03  & 0.02  & 0.03  & 0.02  & 0.04  & 0.02  & 0.03  & 0.02  & 0.03  & 0.02  & 0.03  & 0.02 \\
    $\g_1$ & 0.09  & 0.06  & 0.09  & 0.06  & 0.09  & 0.05  & 0.09  & 0.06  & 0.08  & 0.05  & 0.09  & 0.06 \\
    $\g_2$ & 0.05  & 0.03  & 0.05  & 0.03  & 0.05  & 0.03  & 0.05  & 0.03  & 0.05  & 0.03  & 0.05  & 0.03 \\
    $\tau_{13}$ & -0.15 & 0.29  & -0.10 & 0.28  & -0.24 & 0.28  & -0.05 & 0.27  & -0.20 & 0.26  & -0.09 & 0.31 \\
    $\tau_{23}$ & -0.40 & 0.24  & -0.34 & 0.21  & -0.36 & 0.27  & -0.39 & 0.22  & -0.39 & 0.22  & -0.36 & 0.27 \\\hline
    AIC   & \multicolumn{2}{c}{185.38} & \multicolumn{2}{c}{185.89} & \multicolumn{2}{c}{185.81} & \multicolumn{2}{c}{184.90} & \multicolumn{2}{c}{184.37} & \multicolumn{2}{c}{186.35} \\\hline
    \multicolumn{13}{c}{Vuong's test} \\\hline
  $z_0$  & \multicolumn{2}{c}{3.49} & \multicolumn{2}{c}{2.82} & \multicolumn{2}{c}{1.45} & \multicolumn{2}{c}{3.14} & \multicolumn{2}{c}{7.35} & \multicolumn{2}{c}{1.46} \\
    $p$-value & \multicolumn{2}{c}{$<0.001$} & \multicolumn{2}{c}{0.01} & \multicolumn{2}{c}{0.15} & \multicolumn{2}{c}{$<0.001$} & \multicolumn{2}{c}{$<0.001$} & \multicolumn{2}{c}{0.14} \\
    \hline
    \end{tabular}%
  \label{tab:addlabel}%
\end{table}
\end{landscape}

The AICs showed that a vine copula mixed model with rotated Clayton copulas by 90  and 270 (both for negative dependence but with different tail direction) degrees, beta margins, and permutation $\{13,23,12|3\}$ provides the best fit (Table \ref{pars-perm3-glucan}). This result suggests some skewness to upper-lower tail for the pair of sensitivity   and disease prevalence (i.e., the data display dependence between extreme big values of sensitivity and extreme small values of disease prevalence), and, some skewness to lower-upper tail for the pair of specificity and disease prevalence (i.e., the data display dependence between extreme small values of specificity and extreme big values of disease prevalence). 

It is revealed that a vine copula mixed model with the sensitivity, specificity, and prevalence on the original scale provides better fit than the GLMM, which models the sensitivity, specificity and prevalence on a transformed scale. In fact,  from the Vuong's statistic there is enough improvement to get a highly statistical significant difference ($p$-value$< 0.001$). In order to include a penalty for the different number of parameters among the models, we used the adjusted Vuong's\cite{vuong1989} version: $$\bar D-N^{-1}\{\#\mbox{Model}\,\, 2 \,\,\mbox{parameters}-\#\mbox{Model} \,\,1 \,\, \mbox{parameters}\}.$$

Finally, note that all models roughly agree on the estimated specificity $\hat\pi_1$ and prevalence $\hat\pi_3$, but the estimate $\hat\pi_1$  of sensitivity is smaller when beta margins are assumed.

\subsection{The oral glucose tolerance data}

Ye {\it et al.}\cite{ye-etal-2012}  meta-analyzed 10 studies on the oral glucose tolerance test for the diagnosis of diabetes mellitus in patients during acute coronary syndrome hospitalization.

Fitting the vine copula mixed model for all different permutations,  choices of parametric families of copulas and margins,  the resultant estimate of the one of the  dependence parameter in the first tree  was close to the right boundary of its parameter space, and the conditional dependence parameter was strongly negative. This is a clear case in which  the vine copula model with a full structure  provides more dependence structure that it is actually required. 
This has not been revealed in  Hoyer and Kuss \cite{hoyer&kuss2015} since their approximation  method substantially underestimates the dependence parameters.\cite{nikoloulopoulos2015remada}

Hence we fit  
all the truncated at level-1 vines copula mixed models with different permutations and  choices of parametric families of copulas and margins. 
All models agree on the estimated specificity $\hat\pi_2$ and prevalence $\hat\pi_3$ but the estimate of  $\hat\pi_1$ is larger under the standard GLMM.  Since the number of parameters is the same between the models, we  use the  log-likelihood at  estimates as a  measure for goodness of fit between the  models. According to the likelihood principle better fits are provided using permutation $\{12,13,23|1\}$ and the best is provided by the truncated at level-1 vine copulas composed with BVN copulas and beta margins (Table \ref{perm1-ogt}). The log-likelihood  is  $-86.24$ and $-85.89$ for BVN 
{\color{white} building}

\begin{landscape}
\begin{table}[htbp]
  \centering
  \caption{\label{perm1-ogt}Maximised  log-likelihoods, estimates and standard errors (SE), along with the Vuong's statistics and $p$-values for the oral glucose tolerance data where the dependence is driven by the number of true positives $Y_1$. Vines have been truncated at the 1st level/tree, i.e. $\tau_{23|1}$ is fixed to 0 (conditional independence). }
    \begin{tabular}{ccccccccccccc}
    \hline
    \multicolumn{13}{c}{Normal margins} \\
    \hline
          & \multicolumn{2}{c}{BVN} & \multicolumn{2}{c}{Frank} & \multicolumn{2}{c}{Clayton} & \multicolumn{2}{c}{Clayton by 180} & \multicolumn{2}{c}{Clayton by 0/180} & \multicolumn{2}{c}{Clayton by 180/0} \\
          & Est.  & SE    & Est.  & SE    & Est.  & SE    & Est.  & SE    & Est.  & SE    & Est.  & SE \\\hline
    $\pi_1$ & 0.71  & 0.07  & 0.72  & 0.06  & 0.72  & 0.06  & 0.71  & 0.06  & 0.72  & 0.06  & 0.70  & 0.07 \\
    $\pi_2$ & 0.87  & 0.03  & 0.87  & 0.03  & 0.87  & 0.03  & 0.87  & 0.03  & 0.87  & 0.03  & 0.87  & 0.03 \\
    $\pi_3$ & 0.15  & 0.04  & 0.15  & 0.04  & 0.15  & 0.04  & 0.15  & 0.04  & 0.15  & 0.04  & 0.15  & 0.04 \\
    $\s_1$ & 0.77  & 0.39  & 0.63  & 0.31  & 0.70  & 0.39  & 0.72  & 0.37  & 0.64  & 0.32  & 0.78  & 0.37 \\
    $\s_2$ & 0.67  & 0.20  & 0.69  & 0.22  & 0.71  & 0.23  & 0.64  & 0.20  & 0.71  & 0.24  & 0.63  & 0.19 \\
    $\s_3$ & 0.82  & 0.23  & 0.80  & 0.22  & 0.82  & 0.24  & 0.80  & 0.22  & 0.80  & 0.22  & 0.84  & 0.24 \\
    $\tau_{12}$ & 0.60  & 0.33  & 0.86  & 0.00  & 0.68  & 0.52  & 0.57  & 0.40  & 0.89  & 0.45  & 0.47  & 0.36 \\
    $\tau_{13}$ & 0.19  & 0.31  & 0.01  & 0.23  & 0.09  & 0.29  & 0.14  & 0.27  & 0.06  & 0.22  & 0.24  & 0.30 \\\hline
    $\log L$ & \multicolumn{2}{c}{-86.24} & \multicolumn{2}{c}{-86.49} & \multicolumn{2}{c}{-86.38} & \multicolumn{2}{c}{-86.55} & \multicolumn{2}{c}{-86.36} & \multicolumn{2}{c}{-86.44} \\\hline
    \multicolumn{13}{c}{Vuong's test} \\\hline
   $z_0$ & \multicolumn{2}{c}{-}  & \multicolumn{2}{c}{-0.24} & \multicolumn{2}{c}{-0.23} &  \multicolumn{2}{c}{-0.55} & \multicolumn{2}{c}{-0.13} &  \multicolumn{2}{c}{-0.36} \\
    $p$-value & \multicolumn{2}{c}{-}  & \multicolumn{2}{c}{0.81}  & \multicolumn{2}{c}{0.82}  & \multicolumn{2}{c}{0.58 } & \multicolumn{2}{c}{0.90}  & \multicolumn{2}{c}{0.72} \\\hline
    \multicolumn{13}{c}{Beta margins} \\\hline
          & \multicolumn{2}{c}{BVN} & \multicolumn{2}{c}{Frank} & \multicolumn{2}{c}{Clayton} & \multicolumn{2}{c}{Clayton by 180} & \multicolumn{2}{c}{Clayton by 0/180} & \multicolumn{2}{c}{Clayton by 180/0} \\
          & Est.  & SE    & Est.  & SE    & Est.  & SE    & Est.  & SE    & Est.  & SE    & Est.  & SE \\\hline
    $\pi_1$ & 0.69  & 0.07  & 0.69  & 0.06  & 0.69  & 0.06  & 0.69  & 0.06  & 0.70  & 0.06  & 0.68  & 0.07 \\
    $\pi_2$ & 0.85  & 0.03  & 0.85  & 0.03  & 0.85  & 0.03  & 0.85  & 0.03  & 0.85  & 0.03  & 0.85  & 0.03 \\
    $\pi_3$ & 0.18  & 0.04  & 0.18  & 0.04  & 0.18  & 0.04  & 0.18  & 0.04  & 0.18  & 0.04  & 0.17  & 0.04 \\
    $\g_1$ & 0.11  & 0.09  & 0.10  & 0.09  & 0.10  & 0.08  & 0.10  & 0.09  & 0.09  & 0.07  & 0.11  & 0.09 \\
    $\g_2$ & 0.05  & 0.03  & 0.05  & 0.03  & 0.06  & 0.03  & 0.04  & 0.03  & 0.06  & 0.03  & 0.04  & 0.03 \\
    $\g_3$ & 0.07  & 0.04  & 0.07  & 0.04  & 0.08  & 0.04  & 0.07  & 0.04  & 0.07  & 0.04  & 0.08  & 0.04 \\
    $\tau_{12}$ & 0.56  & 0.29  & 0.63  & 0.40  & 0.60  & 0.30  & 0.52  & 0.36  & 0.73  & 0.41  & 0.43  & 0.35 \\
    $\tau_{13}$ & 0.21  & 0.29  & 0.11  & 0.34  & 0.14  & 0.26  & 0.15  & 0.28  & 0.08  & 0.22  & 0.26  & 0.30 \\\hline
    $\log L$ & \multicolumn{2}{c}{-85.89} & \multicolumn{2}{c}{-86.20} & \multicolumn{2}{c}{-86.00} & \multicolumn{2}{c}{-86.25} & \multicolumn{2}{c}{-86.05} & \multicolumn{2}{c}{-86.09} \\\hline
    \multicolumn{13}{c}{Vuong's test} \\\hline
    $z_0$ & \multicolumn{2}{c}{0.56}  & \multicolumn{2}{c}{ 0.05}  & \multicolumn{2}{c}{ 0.30}  & \multicolumn{2}{c}{ -0.01} & \multicolumn{2}{c}{ 0.19}  & \multicolumn{2}{c}{ 0.20} \\
    $p$-value & \multicolumn{2}{c}{0.57}  & \multicolumn{2}{c}{0.96}  & \multicolumn{2}{c}{0.76}  & \multicolumn{2}{c}{0.99}  & \multicolumn{2}{c}{0.85}  & \multicolumn{2}{c}{0.85} \\
    \hline
    \end{tabular}%
  \label{tab:addlabel}%
\end{table}
\end{landscape}

\noindent building block with  normal and beta margins, respectively;  thus a beta margin seems to be a better fit for the data. However, according to the Vuong's test the the truncated at level-1 vine copula mixed model with beta margins does not provides better fit ($p$-value$=0.574$) than a model with normal margins. The latter is close the the standard GLMM. 

\section{\label{sec-discussion}Discussion}
We have proposed a vine copula mixed model for trivariate  meta-analysis of diagnostic test accuracy studies accounting for disease prevalence. Our general statistical model allows for selection of a bivariate block independently among a variety of parametric copula families, i.e.,
there are no constraints in the choices of bivariate parametric families of copulas. 

It also  includes the trivariate GLMM as a special case  and  can be seen to provide an improvement over the GLMM   on the basis of the log-likelihood and Vuong's \cite{vuong1989} statistic.  Hence,  superior statistical inference for the meta-analytic parameters of interest can be achieved.
This improvement relies on the
fact that the random effects distribution is expressed via copulas which can provide a wide range of flexible  tail dependencies and asymmetries.\cite{joeetal10} 
The MVN assumption is used in statistics because it is mathematically convenient, but one of our points is that the normality assumption is not   reasonable if there is some skewness in the joint tails. The theory and application of copulas have become important in finance, insurance and other areas, in order to deal with dependence in the joint tails. Here, we have indicated that this can also be important in meta-analysis of diagnostic test accuracy studies.

It has been reported in the literature that since the trivariate  GLMM has three correlation parameters, there might   be  estimation problems relating to the correlation parameters, particularly if the sample size is small.\cite{chu-etal-2009} Here we use the notion of ``truncated at level-1 vine copulas", which allow both parsimony and flexible dependence structure, to subside this problem. This is  due to the main result in Joe {\it et al.}\cite{joeetal10}:  
all bivariate margins of the vine copula have upper/lower
tail dependence if the bivariate copulas at level 1 have upper/lower
tail dependence. In a copula modelling framework as well as in any other statistical modelling framework, the pursuit of perfection is illusory and a balance should always be struck between fit and parsimony.\cite{genest&favre06}

A challenging  direction of future research is to extend our framework  to model data from studies that  do not report all  outcomes via pattern mixture models. 
Pattern mixture models
are studied in Shen and Weissfeld\cite{shen-Weissfeld06} for multivariate copulas and in Mavridis {\it et al.}\cite{mavridis-etal-2014}   for  multivariate meta-analysis.

\section*{\label{software}Software}
{\tt R} functions to  implement the vine copula mixed model for meta-analysis of diagnostic test accuracy studies acounting for disease prevalence are  part of the {\tt  R} package {\tt  CopulaREMADA}. \cite{Nikoloulopoulos-2015}


\bibliographystyle{vancouver}



\end{document}